\documentclass[11pt,a4paper]{article}
\pdfoutput=1

\usepackage{jcappub}
\usepackage{multirow}
\usepackage{amssymb} 
\usepackage{amsmath}
\usepackage[normalem]{ulem}
\usepackage{subfigure}

\newcommand{\HL}{{H_{\rm L}}}
\newcommand{\HT}{{H_{\rm T}}}
\newcommand{\nab}{\nabla}

\begin{document}
	
	\title{The large-scale general-relativistic correction for Newtonian mocks}
	
	\date{\today}
	
	\author[a]{Julian Adamek}
	\emailAdd{julian.adamek@qmul.ac.uk}
	
	\author[b]{and Christian Fidler}
	\emailAdd{fidler@physik.rwth-aachen.de}
	
	\affiliation[a]{School of Physics \& Astronomy, Queen Mary University of London, 327 Mile End Road, London E1 4NS, UK}

	\affiliation[b]{Institute for Theoretical Particle Physics and Cosmology (TTK), RWTH Aachen University, Otto-Blumenthal-Strasse, D--52057 Aachen, Germany}

	\abstract{We clarify the subtle issue of finding the correct mapping of Newtonian simulations to light-cone observables at very large distance scales. A faithful general-relativistic interpretation specifies a gauge, i.e.\ a chart that relates the simulation data to points of the space-time manifold. It has already been pointed out that the implicit gauge choice of Newtonian simulations is indeed different from the Poisson gauge that is commonly adopted for relativistic calculations, the difference being most significant at large scales. It is therefore inconsistent, for example, to predict weak-lensing observables from simulations unless this gauge issue is properly accounted for. Using perturbation theory as well as fully relativistic N-body simulations we quantify the systematic error introduced this way, and we discuss several solutions that would render the calculations relativistically self-consistent.

\bigskip
{\noindent\footnotesize This is the Accepted Manuscript version of an article accepted for publication in Journal of Cosmology and Astroparticle Physics. Neither SISSA Medialab Srl nor IOP Publishing Ltd is responsible for any errors or omissions in this version of the manuscript or any version derived from it. The Version of Record is available online at \url{https://doi.org/10.1088/1475-7516/2019/09/026}.}	
	}
	
	\maketitle   
	
	\flushbottom
	\section{Introduction}
	\label{sec:intro}
	
	N-body simulations play an important role in studying the large-scale structure of the Universe because they facilitate accurate non-perturbative predictions. The vast majority of these simulations work in the Newtonian limit in which the geometric nature of gravity is not made explicit. Bypassing the conceptual overhead of general relativity one can do simplified analyses that work well on small scales, but
	relativistic effects become important on the larger scales especially when approaching the
	cosmological horizon. As astronomical surveys become larger and deeper one therefore has to rethink how one can maintain accurate predictions from N-body simulations.
	
	The correspondence between Newtonian and relativistic dynamics has been discussed since the advent of general relativity \cite{Einstein:1916vd}, but has attracted renewed interest due to cosmological implications \cite{Chisari:2011iq,Green:2011wc,Flender:2012nq}. It has been pointed out in \cite{Chisari:2011iq} that the spatial coordinates in Newtonian simulations are not compatible with the Poisson gauge at large scales, and a coordinate remapping was suggested and quantified. This principle has since been put on a firm relativistic footing by understanding the coordinate remapping as a coordinate transformation from the so-called N-body gauge introduced in \cite{Fidler:2015npa}. Generalising this idea led to the development of a more versatile class of gauges, so-called Newtonian motion gauges, that can absorb additional large-scale relativistic effects such as the ones of radiation \cite{Adamek:2017grt} or neutrinos \cite{Fidler:2018bkg}. This framework provides a precise understanding how the gauge-dependent simulation data can be mapped to observables. Since astronomical observations are almost exclusively based on electromagnetic signals this mapping involves the characterisation of null geodesics. One can proceed consistently in a gauge adapted to a Newtonian simulation but the analysis then differs from the traditional ones which are usually based on Poisson gauge (sometimes implicitly) \cite{Adamek:2017kir,NLNM}. In other words, it is inconsistent to use the lensing formula in Poisson gauge naively to generate mock observables from a Newtonian simulation. The main aim of this paper is to quantify this issue more precisely.
	
	One possible approach is to compute null geodesics consistently in a Newtonian motion gauge, as we will discuss in some detail towards the end of this article. There are, however, several other solutions to address this systematic error. The issue can be avoided entirely, of course, if one employs relativistic simulations that are directly framed in the Poisson gauge \cite{Adamek:2015eda,Adamek:2016zes}. Alternatively one can translate a Newtonian simulation \textit{a posteriori} into Poisson gauge before
	proceeding with the analysis, as we illustrate below.
	
	It should be clear from the outset that our discussion is only relevant in the context of a relativistic interpretation of Newtonian simulation data. The well-known results from cosmological perturbation theory \cite{Kasai:1987ap,Yoo:2009au,Bonvin:2011bg,Challinor:2011bk} that were obtained through fully relativistic calculations stand unchallenged. This also holds for some perturbative arguments that relate to linear Newtonian solutions \cite{Haugg:2012ng}. However, the main point of N-body simulations is to probe into the non-perturbative regime of matter evolution. We argue that a consistent relativistic interpretation can still be achieved in this regime, relying on a weak-field description of gravity alone. In particular, such an interpretation should not refer to fluid variables as this would be a poor description of matter in the non-perturbative regime. For instance, there is no unique notion of a velocity field in the real Universe, and thus our construction of observables should not depend on one.
	
	This paper is organised as follows. We first summarise the concept behind the N-boisson gauge, which provides a rigorous mathematical description of the problem, in Section \ref{sec:Nbgauge}. In Section \ref{sec:observables} we quantify the impact on observables using a large relativistic N-body simulation. We conclude in Section \ref{sec:conclusions}.
	
	\section{The N-boisson gauge}
	\label{sec:Nbgauge}
	
	The N-boisson gauge is a specific limit of the class of Newtonian motion gauges \cite{Fidler:2018bkg}. In these the relativistic corrections to the geodesics of dark matter particles are absorbed in a particular choice of coordinates. The dark matter particles then follow Newtonian equations of motion that are interpreted within the coordinates of the underlying gauge. 
	
	In the case of a Universe that is filled with only cold dark matter and a cosmological constant the N-boisson gauge is a particularly simple solution to this problem. 
	We define a yet unfixed gauge by the metric  
	\begin{subequations}
    \label{metric-potentials}
    \begin{align}
    g_{00} &= -a^2 \left[ 1 + 2 A \right] \,, \\
    g_{0i} &=  -a^2  \nab_i  B  \,,\\
    g_{ij} &= a^2 \left[ \delta_{ij} \left( 1 + 2 \HL \right) - 2 \left(  \nab_i \nab_j - \frac {\delta_{ij}}  {3} \Delta\right) \HT  \right] \,.
    \end{align}
    \end{subequations}
    with the conformal time $\tau$ and the scale factor $a=a(\tau)$, considering only scalar perturbations.
	
	The N-boisson gauge is specified by requiring that $\Delta\HT = 3\zeta$ with the comoving curvature perturbation $\zeta$, which is constant at late times in a standard cosmology. 
	The temporal gauge is fixed by connecting the shift vector $B$ to the time derivative of $\HT$ such that to high accuracy $B \approx 0$. The N-boisson gauge shares the temporal gauge condition with the Poisson gauge (which has $\HT = 0, B= 0$), while the spatial coordinates are displaced according to the curvature perturbation. It was further shown that, in a realistic Universe that contains some radiation initially, Newtonian motion gauges can be found which converge towards the N-boisson gauge at late times. In particular this limit is realised when using the popular backscaling initial conditions \cite{Fidler:2017ebh}. Therefore most of the Newtonian N-body simulations can be interpreted in the N-boisson gauge and are indeed implicitly run in the corresponding coordinates.
	
	This has important implications for the propagation of light rays. In the coordinates used by the simulations
	the presence of a non-vanishing $\HT$ introduces an extra contribution along the line of sight. The correction can be described by a potential, depending on the difference in $\HT$ between the point of emission and absorption. It is similar to the Sachs-Wolfe effect, describing the energy-loss of photons that escape a gravitational potential. Only that instead of changing the energy, the metric potential $\HT$ changes the angle between emission and absorption and therefore shifts the observed galaxies around on the sky \cite{NLNM}.
	
	Since $\HT$ only varies on larger scales and is locally almost constant this integrated coordinate shift (ICS)
	does not impact local observations on small scales, as it displaces local structures coherently. However, when looking at more distant objects there is a well understood angular displacement that can be accounted for with little computational effort. 
	
	The simplest way to do this is by spatially displacing the simulated galaxies to their Poisson-gauge positions by a spatial coordinate transformation $L^i = \nab^i \HT$. 
	This transformation can be derived directly from the comoving curvature perturbation, since the Poisson and N-boisson gauge already share their temporal gauge condition.
	It is important to note that the lensing potential satisfies different relations to the density in Poisson gauge and N-boisson gauge. A good approximation is obtained by computing it from the density in N-boisson gauge (where the relation is simpler) but it then still needs to be displaced according to the coordinate transformation in order to match up with the non-linear structure.
	
	\section{The impact on observables}
	\label{sec:observables}

    We quantify the impact of the ICS using a large N-body simulation carried out with the relativistic code \textit{gevolution}. In order to include the large-scale effects we are interested in we use a simulation box of 9.6 Gpc$/h$ with 7680$^3$ particles. This gives a mass resolution of about 1.7$\times$10$^{11} M_\odot/h$. We construct a full-sky light cone of particles, i.e.\ we fix an observer position (the center of the box) and record the position and peculiar velocity of each particle as its world-line crosses the past light cone of the observer. Note that these are directly given in Poisson gauge within our relativistic simulation framework. The light cone goes to a maximum distance of 4.8 Gpc$/h$ (corresponding to redshift $z_\mathrm{max} \simeq 3.6$) in order to avoid issues related to the finite size of the simulation box.
    
    In a post-processing step we run the \textit{ROCKSTAR} halo finder \cite{Behroozi:2011ju} on the particle data and thus obtain a halo catalog on the light cone with some 1.6 million halos above 3.3$\times$10$^{13} M_\odot/h$. Smaller structure is not well resolved in our simulation, but we are interested here in the large-scale distribution of halos. From the same pseudo-random number sequence that was used to prepare the initial data, we also generate a
    realisation of the curvature perturbation using the linear Einstein-Boltzmann code \textit{CLASS} \cite{Blas:2011rf}. This allows us to compute the coordinate transformation between Poisson and N-boisson gauge, i.e.\ we can shift the halos back to the positions they would have in a Newtonian simulation of the same realisation. The high accuracy of this method has been established in \cite{Adamek:2017grt} (see Section 4.3 therein) and spares us from having to re-run the full simulation in N-boisson gauge.
    
    \begin{figure}[t!]
    \centerline{\includegraphics[width=0.85\textwidth]{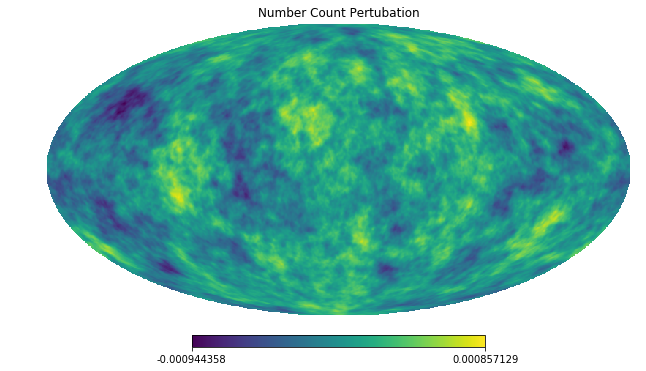}}
    \caption{We show the number-count perturbation induced by the displacement when applied to a homogeneous field, for a redshift bin at $z=1.5$ with a width of $100$ Mpc$/h$.}
    \label{numbercountmap}
    \end{figure}
    
    We translate both ``versions'' of the catalog (Poisson gauge and N-boisson gauge) to redshift space, accounting only for Doppler redshift-space distortions (RSD) for simplicity. Additional relativistic corrections like lensing are not included here since we assume that these would be computed in a further post-processing step identically in both cases, but only consistent with halo positions given in Poisson gauge. Therefore
    the difference between both
    halo distributions in redshift space is precisely due to
    the ICS effect.
    
    We analyse the number counts and their angular power spectra $C_\ell$ for different window functions in redshift.
    While we present the
    ICS effect
    here from the point of view of relativistic light propagation, it should be noted that it is not possible to distinguish between effects on the light propagation or structure evolution in isolation as this distinction is gauge-dependent. The ICS effect is the missing relativistic correction that is usually neglected in the analysis of large-scale structure simulations and including it provides the full result that has both a consistent relativistic particle evolution and relativistic ray-tracing equations. However, for the case of N-boisson gauge, the latter are not widely known in the community and we will therefore elaborate on them below. We also refer the reader to Section 5 of \cite{NLNM} for further details.

    \begin{figure}[tpb]
    \centerline{\includegraphics[width=0.8\textwidth]{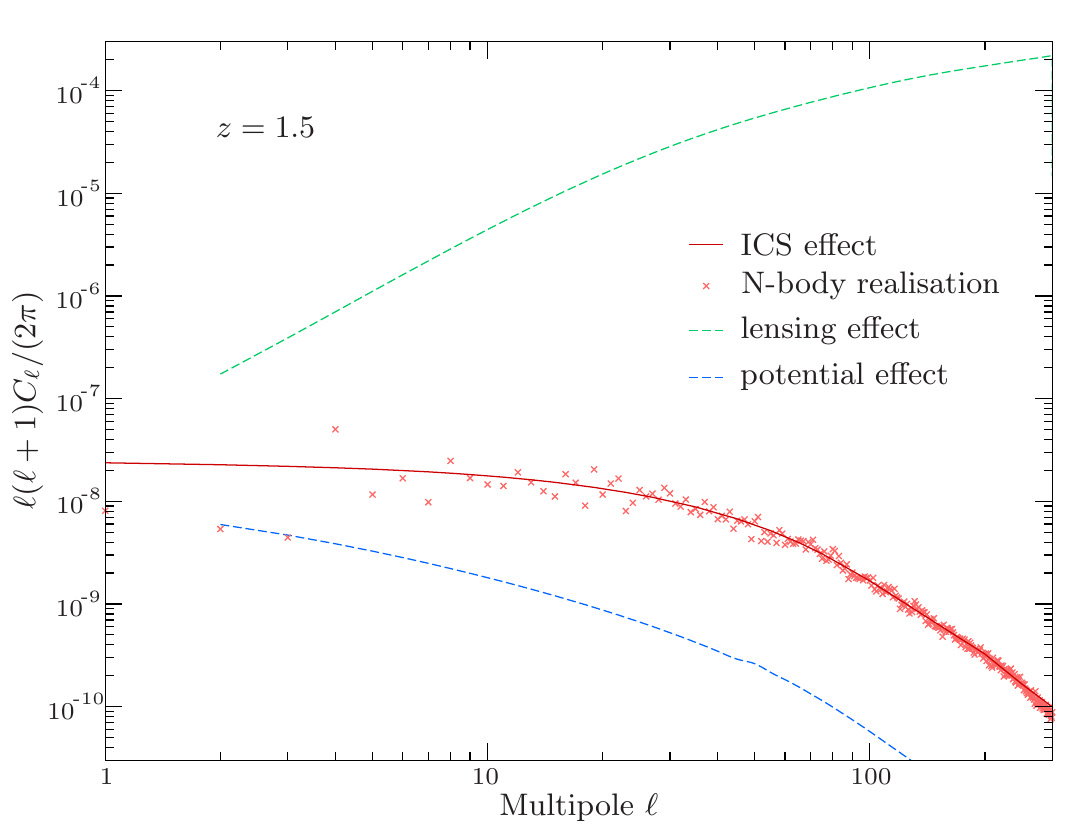}}
    \caption{We show the number-count perturbation power $C_\ell$ induced by the integrated coordinate shift (ICS) in a redshift bin at $z=1.5$ with a width of $100$ Mpc$/h$, as a function of multipole $\ell$ taken on the observer's sky. The crosses mark the particular realisation of our large N-body simulation (shown in Figure \ref{numbercountmap}), whereas the solid red line is the prediction obtained from Eq.~(\ref{Clprediction}). The green dashed line shows the number-count perturbation induced by weak gravitational lensing alone, as predicted by \textit{CLASS} for the same redshift bin. Note that the ICS and lensing effect are correlated, which is not studied here. The lower, blue dashed line shows the additional relativistic contributions (due to the gravitational potential) as predicted by \textit{CLASS}.}
    \label{densityCl}
    \end{figure}

    The ICS is described by a large-scale displacement field $x^i \rightarrow x^i + L^i$ that is applied to the galaxy distributions that have power on much smaller scales. It creates two distinct signatures. First, by displacing a nearly homogeneous background of galaxies, the structure of the displacement field is imprinted in the data. This is illustrated in Figure \ref{numbercountmap} and is responsible for a large-scale modification of the power spectrum. 
    Second, when acting on the inhomogeneous galaxy over-densities it modulates the correlation of small-scale density perturbations. The ICS appears in addition to the other relativistic corrections precisely because of employing the densities and velocities extracted from a Newtonian N-body simulation.
    
    \begin{figure}[ht]
    \centerline{\includegraphics[width=.85\textwidth]{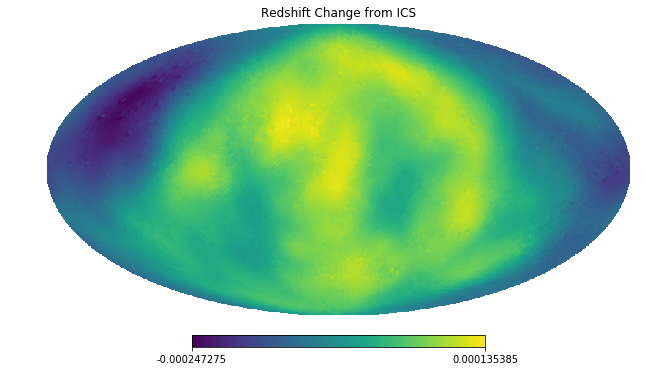}}
    \centerline{\includegraphics[width=.85\textwidth]{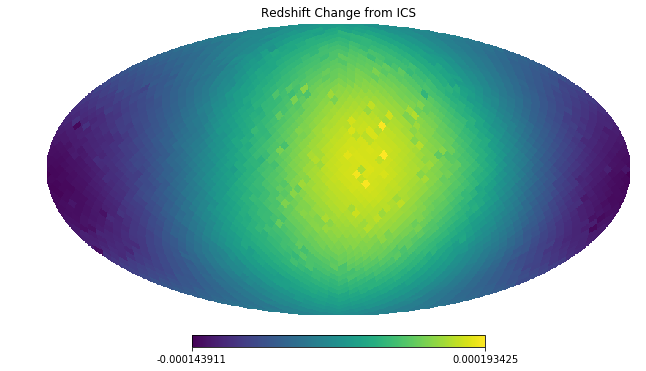}}
    \caption{Change in observed redshift averaged for halos within a pixel and redshift bin due to the ICS. The top panel covers a redshift bin of $1.35 < z < 1.65$, while the bottom panel is using a much more local measurement $0.1 < z < 0.3$. Pixels are chosen such that they include a sufficient number of halos. The ICS is inducing a large-scale redshift distortion.}
    \label{dispz}
    \end{figure}
    
    \begin{figure}[t]
    \centerline{\includegraphics[width=.85\textwidth]{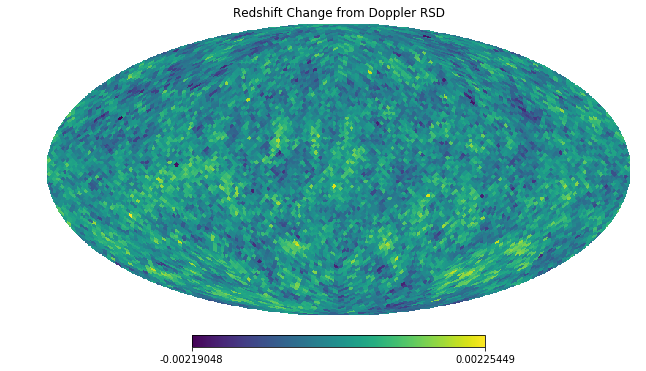}}
    \centerline{\includegraphics[width=.85\textwidth]{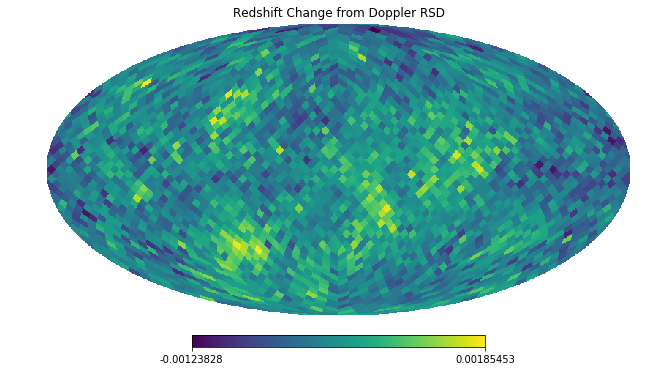}}
    \caption{Change in observed redshift from Doppler RSD for the same redshift bins used in Figure \ref{dispz}. Compared to the ICS the Doppler terms are not coherent over larger scales. The Doppler RSD is roughly ten times larger than the redshift distortion due to the ICS.}
    \label{dispzdoppler}
    \end{figure}
    
    As is evident from Figure \ref{numbercountmap} the number-count perturbation induced by the displacement is correlated over several degrees (at redshift $z=1.5$) and typically of the order of $\text{few}\times 10^{-4}$ (note that the amplitude is independent of redshift). In order to get a signal-to-noise larger than unity one therefore needs some ten million galaxies, more than we have halos in our catalog. However, galaxy mocks well in excess of that number are routinely used to set up the analysis pipelines for current and upcoming surveys, and we therefore expect this effect to be relevant in some cases.
    
    To quantify this further we show in Figure \ref{densityCl} the angular power spectrum for the large-scale effect of the displacement. As in Figure \ref{numbercountmap} we choose a redshift bin at $z = 1.5$ with a radial extent of $100$ Mpc$/h$ and compare the induced number-count perturbation with the one expected from lensing (green dashed) and subleading relativistic corrections (blue dashed). We use the
    public Einstein-Boltzmann code \textit{CLASS}, in particular the methods of \cite{DiDio:2013bqa}, for the latter two predictions. For instance, the lensing contribution is sourced by the term $\Delta_\ell^\mathrm{Len}$ as defined in Eq.~(A.14) of that reference.
    
    While \textit{CLASS} does not readily compute the ICS effect (it originates from an unusual gauge choice -- the one that is implicitly assumed in Newtonian N-body simulations) it can still be predicted from \textit{CLASS} output. The large-scale number counts receive a contribution $-\nabla_i L^i = -\Delta \HT = -3\zeta$ from the displacement. A quick calculation then shows that the angular correlation of this contribution is given by
    \begin{equation}
    \label{Clprediction}
        C_\ell^\mathrm{ICS} = 4\pi \int_0^\infty \frac{dk}{k}9 \Delta^\zeta(k) \left[\int_0^\infty W(r) j_{\ell}(k r) dr\right]^2\,,
    \end{equation}
    where $W(r)$ is the (normalised) radial window function and $\Delta^\zeta(k)$ is the power spectrum of $\zeta$. The latter can be obtained directly from the transfer functions that \textit{CLASS} provides.
    
    The solid red curve in Figure \ref{densityCl} shows $C_\ell^\mathrm{ICS}$ for the relevant tophat window function.
    Interestingly, in terms of amplitude (the square-root of the power) the ICS effect is only about one order of magnitude smaller than the lensing on the first ten multipoles, and it dominates over the remaining relativistic corrections (which are related to the gravitational potential) on all angular scales. Note that this comparison only illustrates the linear large-scale impact of the ICS due to a displacement of a homogeneous sample of galaxies. In reality the displacement acts on an inhomogeneous distribution and additionally transfers power between different multipoles.
    
    \begin{figure}[t]
    \centerline{\includegraphics[width=.85\textwidth]{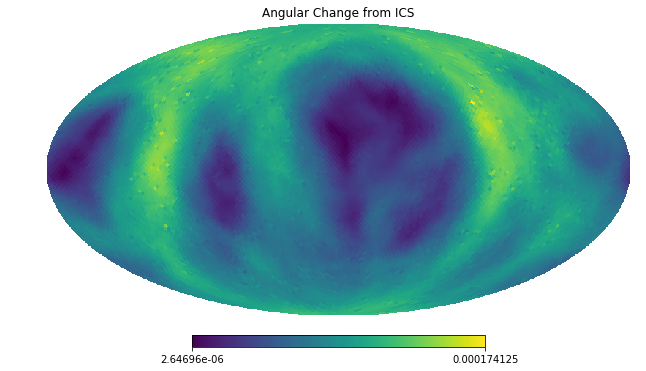}}
    \centerline{\includegraphics[width=.85\textwidth]{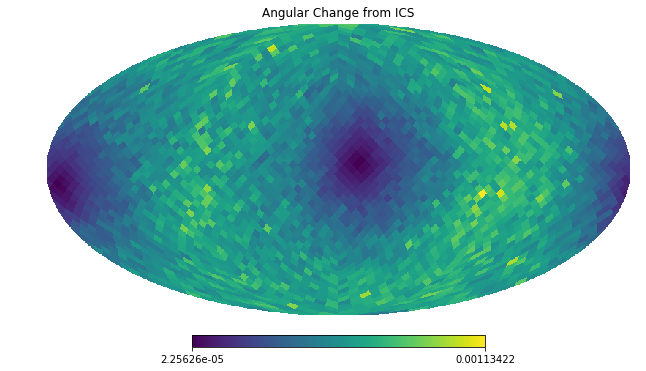}}
    \caption{Averaged angular displacement (in radians) in the observed sky positions for the halos within the redshift bins used for Figures \ref{dispz} and \ref{dispzdoppler}. Again we find a large scale coherent displacement of the observed structures.}
    \label{dispr}
    \end{figure}
    
    In addition to the number counts, the ICS affects both the observed redshift and the angular position of the galaxies on the sky. The changes to the redshift are typically of the order of $10^{-4}$ shown in Figure \ref{dispz}, and can be understood as a result of a change in coordinate distance between the two gauges (note that we consistently displace the observer to avoid artefacts: all effects discussed here are the result of \textit{relative} displacements).
    For comparison we show the redshift distortions induced from the well-known Doppler terms for the same halos in Figure \ref{dispzdoppler}. The contribution of ICS is about 10\% relative to those. We note that while the Doppler RSD is randomly scattered across the sky, the ICS is coherent on degree scales. It becomes particularly simple when observing local structures, for which the curvature perturbation at emission is still significantly correlated with the one at absorption.

    Figure \ref{dispr} shows the angular displacement (the absolute value of the displacement angle) on the observer's sky from the ICS. We find that objects are shifted by up to a few arcminutes, coherently over several degrees. Given a sufficiently large number-density of sources this displacement will lead to number-count perturbations as illustrated in Figure \ref{numbercountmap}.
    
    The angular displacement due to the ICS can also be worked out by considering weak lensing from the perspective of N-boisson gauge. The photon geodesic equations read
    \begin{equation}
        \frac{k^i}{k^0} = \frac{dx^i}{d\tau} = n^i \left(1 + A - \HL\right) + n^j \left(\nabla_j\nabla^i - \frac{\delta^i_j}{3} \Delta\right)\HT\,,
    \end{equation}
    \begin{equation}
        \frac{dn^i}{d\tau} = -\left(\delta^{ij} - n^i n^j\right) \nabla_j \left(A - \HL - \frac{1}{3}\Delta \HT\right)\,,
    \end{equation}
    where $n^i$ is the direction of propagation ($\delta_{ij} n^i n^j = 1$) and we used that $B \simeq 0$ in this gauge. Note also that due to the gauge condition $\HT' = 0$ the combination $A - \HL - \Delta\HT/3 = 2 \Psi_\mathrm{W}$, where $\Psi_\mathrm{W}$ denotes the Weyl potential (the mean of the two Bardeen potentials). A first-order integral of these equation yields a deflection angle of
    \begin{equation}
        \delta\theta^i = -\frac{\delta^{ij} - n^i n^j}{\chi_\mathrm{s}} \left[ \bigl.\nabla_j\HT\bigr\vert_\mathrm{o} - \bigl.\nabla_j\HT\bigr\vert_\mathrm{s} + 2 \int_0^{\chi_\mathrm{s}} \left(\chi_\mathrm{s} - \chi\right) \nabla_j\Psi_\mathrm{W} d\chi\right]\,,
    \end{equation}
    where $\chi_\mathrm{s}$ denotes the conformal distance to the source. It is common practice to introduce a lensing potential $\psi$ on the celestial sphere such that the deflection angle $\delta\theta^i$ is the gradient of $\psi$ on the sphere. In Poisson gauge, where $\HT = 0$, the lensing potential is
    \begin{equation}
        \psi^\mathrm{P} = -2 \int_0^{\chi_\mathrm{s}} \frac{\chi_\mathrm{s} - \chi}{\chi_\mathrm{s} \chi} \Psi_\mathrm{W} d\chi\,,
    \end{equation}
    whereas in N-boisson gauge the lensing potential acquires two additional terms,
    \begin{equation}
        \psi^\mathrm{Nb} = \psi^\mathrm{P} + \frac{\bigl.\HT\bigr\vert_\mathrm{s}}{\chi_\mathrm{s}^2} - \frac{\theta^i \bigl.\nabla_i\HT\bigr\vert_\mathrm{o}}{\chi_\mathrm{s}}\,.
    \end{equation}
    The two additional terms incorporate the parallax from the coordinate shift of source and observer, respectively. This ensures that the observed sky position remains invariant. In other words, one can either choose to apply the lensing correction in Poisson gauge as usual, in which case the positions have to be transformed to that coordinate system first, or one can apply the lensing correction directly on the data in N-boisson gauge, in which case the lensing potential absorbs the transformation.

    \section{Conclusions}
    \label{sec:conclusions}
	
	We have studied the large-scale relativistic correction to the observed galaxy distribution in Newtonian mock catalogs. Ignoring this correction introduces a systematic error in the number counts on angular scales larger than a few degrees, reminiscent of a map distortion that shifts the sky positions by a few arcminutes. Small-scale correlations are also modulated by this distortion. At redshift $z \sim 1.5$ the error dominates over all other relativistic corrections except for lensing, which is however only about one order of magnitude larger on angular scales above $\sim 20$ degrees. We estimate that the relativistic correction becomes statistically significant for wide-angle galaxy mocks with more than ten million objects.
	
	Similarly to the angular positions the observed redshifts of galaxies are also affected, but most surveys have a good resolution of the angle while the resolution on the redshift is rather poor. In this case the impact on the observed redshift can be neglected.
	
	We provide a simple recipe to remove the systematic error that is introduced when neglecting relativistic corrections in ordinary Newtonian N-body simulations. It is sufficient to displace all particles in a post-processing stage and this can be done on snapshots even after the simulation is completed, given that the initial pseudo-random number sequence is available. Alternatively, the displacement can be absorbed into the lensing potential.  
	
	Considering the size of the correction and the sensitivity of future surveys and cosmic variance we conclude that this correction should be included in the analysis of high precision large-scale galaxy mocks. In particular, any endeavours to test general relativity using the subleading light-cone projection effects should only rely on numerical models that properly account for this correction.

	\section*{Acknowledgements}
	
	We thank Sebastian Solibida who has been doing preliminary work for this paper in his Bachelor Thesis. JA acknowledges funding by STFC Consolidated Grant ST/P000592/1. This work was supported by a grant from the Swiss National Supercomputing Centre (CSCS) under project ID s710.

	\bibliography{references}

\providecommand{\href}[2]{#2}\begingroup\raggedright\begin{thebibliography}{10}

\bibitem{Einstein:1916vd}
A.~Einstein, \emph{{The Foundation of the General Theory of Relativity}},
  \href{http://dx.doi.org/10.1002/andp.200590044,
  10.1002/andp.19163540702}{\emph{Annalen Phys.} {\bf 49} (1916) 769--822}.

\bibitem{Chisari:2011iq}
N.~E. Chisari and M.~Zaldarriaga, \emph{{Connection between Newtonian
  simulations and general relativity}},
  \href{http://dx.doi.org/10.1103/PhysRevD.84.089901,
  10.1103/PhysRevD.83.123505}{\emph{Phys. Rev.} {\bf D83} (2011) 123505},
  [\href{http://arxiv.org/abs/1101.3555}{{\tt 1101.3555}}].

\bibitem{Green:2011wc}
S.~R. Green and R.~M. Wald, \emph{{Newtonian and Relativistic Cosmologies}},
  \href{http://dx.doi.org/10.1103/PhysRevD.85.063512}{\emph{Phys. Rev.} {\bf
  D85} (2012) 063512}, [\href{http://arxiv.org/abs/1111.2997}{{\tt
  1111.2997}}].

\bibitem{Flender:2012nq}
S.~F. Flender and D.~J. Schwarz, \emph{{Newtonian versus relativistic
  cosmology}}, \href{http://dx.doi.org/10.1103/PhysRevD.86.063527}{\emph{Phys.
  Rev.} {\bf D86} (2012) 063527}, [\href{http://arxiv.org/abs/1207.2035}{{\tt
  1207.2035}}].

\bibitem{Fidler:2015npa}
C.~Fidler, C.~Rampf, T.~Tram, R.~Crittenden, K.~Koyama and D.~Wands,
  \emph{{General relativistic corrections to $N$-body simulations and the
  Zel'dovich approximation}},
  \href{http://dx.doi.org/10.1103/PhysRevD.92.123517}{\emph{Phys. Rev.} {\bf
  D92} (2015) 123517}, [\href{http://arxiv.org/abs/1505.04756}{{\tt
  1505.04756}}].

\bibitem{Adamek:2017grt}
J.~Adamek, J.~Brandbyge, C.~Fidler, S.~Hannestad, C.~Rampf and T.~Tram,
  \emph{{The effect of early radiation in N-body simulations of cosmic
  structure formation}},
  \href{http://dx.doi.org/10.1093/mnras/stx1157}{\emph{Mon. Not. Roy. Astron.
  Soc.} (2017) }, [\href{http://arxiv.org/abs/1703.08585}{{\tt 1703.08585}}].

\bibitem{Fidler:2018bkg}
C.~Fidler, A.~Kleinjohann, T.~Tram, C.~Rampf and K.~Koyama, \emph{{A new
  approach to cosmological structure formation with massive neutrinos}},
  \href{http://dx.doi.org/10.1088/1475-7516/2019/01/025}{\emph{JCAP} {\bf 1901}
  (2019) 025}, [\href{http://arxiv.org/abs/1807.03701}{{\tt 1807.03701}}].

\bibitem{Adamek:2017kir}
J.~Adamek, \emph{{Perturbed redshifts from N-body simulations}},
  \href{http://dx.doi.org/10.1103/PhysRevD.97.021302}{\emph{Phys. Rev.} {\bf
  D97} (2018) 021302}, [\href{http://arxiv.org/abs/1708.07552}{{\tt
  1708.07552}}].

\bibitem{NLNM}
C.~Fidler, T.~Tram, C.~Rampf, R.~Crittenden, K.~Koyama and D.~Wands,
  \emph{{General relativistic weak-field limit and Newtonian N-body
  simulations}},
  \href{http://dx.doi.org/10.1088/1475-7516/2017/12/022}{\emph{JCAP} {\bf 1712}
  (2017) 022}, [\href{http://arxiv.org/abs/1708.07769}{{\tt 1708.07769}}].

\bibitem{Adamek:2015eda}
J.~Adamek, D.~Daverio, R.~Durrer and M.~Kunz, \emph{{General relativity and
  cosmic structure formation}},
  \href{http://dx.doi.org/10.1038/nphys3673}{\emph{Nature Phys.} {\bf 12}
  (2016) 346--349}, [\href{http://arxiv.org/abs/1509.01699}{{\tt 1509.01699}}].

\bibitem{Adamek:2016zes}
J.~Adamek, D.~Daverio, R.~Durrer and M.~Kunz, \emph{{gevolution: a cosmological
  N-body code based on General Relativity}},
  \href{http://dx.doi.org/10.1088/1475-7516/2016/07/053}{\emph{JCAP} {\bf 1607}
  (2016) 053}, [\href{http://arxiv.org/abs/1604.06065}{{\tt 1604.06065}}].

\bibitem{Kasai:1987ap}
M.~Kasai and M.~Sasaki, \emph{{The Number Count Redshift Relation in a
  Perturbed Friedmann Universe}},
  \href{http://dx.doi.org/10.1142/S0217732387000902}{\emph{Mod. Phys. Lett.}
  {\bf A2} (1987) 727}.

\bibitem{Yoo:2009au}
J.~Yoo, A.~L. Fitzpatrick and M.~Zaldarriaga, \emph{{A New Perspective on
  Galaxy Clustering as a Cosmological Probe: General Relativistic Effects}},
  \href{http://dx.doi.org/10.1103/PhysRevD.80.083514}{\emph{Phys. Rev.} {\bf
  D80} (2009) 083514}, [\href{http://arxiv.org/abs/0907.0707}{{\tt
  0907.0707}}].

\bibitem{Bonvin:2011bg}
C.~Bonvin and R.~Durrer, \emph{{What galaxy surveys really measure}},
  \href{http://dx.doi.org/10.1103/PhysRevD.84.063505}{\emph{Phys. Rev.} {\bf
  D84} (2011) 063505}, [\href{http://arxiv.org/abs/1105.5280}{{\tt
  1105.5280}}].

\bibitem{Challinor:2011bk}
A.~Challinor and A.~Lewis, \emph{{The linear power spectrum of observed source
  number counts}},
  \href{http://dx.doi.org/10.1103/PhysRevD.84.043516}{\emph{Phys. Rev.} {\bf
  D84} (2011) 043516}, [\href{http://arxiv.org/abs/1105.5292}{{\tt
  1105.5292}}].

\bibitem{Haugg:2012ng}
T.~Haugg, S.~Hofmann and M.~Kopp, \emph{{Newtonian N-body simulations are
  compatible with cosmological perturbation theory}},
  \href{http://arxiv.org/abs/1211.0011}{{\tt 1211.0011}}.

\bibitem{Fidler:2017ebh}
C.~Fidler, T.~Tram, C.~Rampf, R.~Crittenden, K.~Koyama and D.~Wands,
  \emph{{Relativistic initial conditions for N-body simulations}},
  \href{http://dx.doi.org/10.1088/1475-7516/2017/06/043}{\emph{JCAP} {\bf 1706}
  (2017) 043}, [\href{http://arxiv.org/abs/1702.03221}{{\tt 1702.03221}}].

\bibitem{Behroozi:2011ju}
P.~S. Behroozi, R.~H. Wechsler and H.-Y. Wu, \emph{{The Rockstar Phase-Space
  Temporal Halo Finder and the Velocity Offsets of Cluster Cores}},
  \href{http://dx.doi.org/10.1088/0004-637X/762/2/109}{\emph{Astrophys. J.}
  {\bf 762} (2013) 109}, [\href{http://arxiv.org/abs/1110.4372}{{\tt
  1110.4372}}].

\bibitem{Blas:2011rf}
D.~Blas, J.~Lesgourgues and T.~Tram, \emph{{The Cosmic Linear Anisotropy
  Solving System (CLASS) II: Approximation schemes}},
  \href{http://dx.doi.org/10.1088/1475-7516/2011/07/034}{\emph{JCAP} {\bf 1107}
  (2011) 034}, [\href{http://arxiv.org/abs/1104.2933}{{\tt 1104.2933}}].

\bibitem{DiDio:2013bqa}
E.~Di~Dio, F.~Montanari, J.~Lesgourgues and R.~Durrer, \emph{{The CLASSgal code
  for Relativistic Cosmological Large Scale Structure}},
  \href{http://dx.doi.org/10.1088/1475-7516/2013/11/044}{\emph{JCAP} {\bf 1311}
  (2013) 044}, [\href{http://arxiv.org/abs/1307.1459}{{\tt 1307.1459}}].

\end{thebibliography}\endgroup
	\bibliographystyle{JHEP}

\end{document}